\documentclass[12pt]{article}

\usepackage[xdvi]{graphicx}
\usepackage{latexsym}
\usepackage{psfrag}
\usepackage{subfigure}
\usepackage{amssymb}

\begin{document}  
\newcommand{\todo}[1]{{\em \small {#1}}\marginpar{$\Longleftarrow$}}  
\newcommand{\labell}[1]{\label{#1}\qquad_{#1}} 
 
\rightline{DCPT-01/67}  
\rightline{hep-th/0111247}  
\vskip 1cm 
\centerline{\Large \bf CFTs on Non-Critical Braneworlds}  
\vskip 1cm  
  
\renewcommand{\thefootnote}{\fnsymbol{footnote}}  
\centerline{\bf  
 Antonio Padilla}
\vskip .5cm  
\centerline{ \it Centre for Particle Theory, Department of 
Mathematical Sciences}  
\centerline{\it University of Durham, South Road, Durham DH1 3LE, U.K.}  
\vskip .5cm
\centerline{Antonio.Padilla@durham.ac.uk}  
\setcounter{footnote}{0}  
\renewcommand{\thefootnote}{\arabic{footnote}}  
 

\def\real{I\negthinspace R}
\def\zed{Z\hskip -3mm Z }
\def\half{\textstyle{1\over2}}
\def\quarter{\textstyle{1\over4}}
\def\sech{\,{\rm sech}\,}
\def\ie{{\it i.e.,}}
\newcommand{\be}{\begin{equation}}
\newcommand{\ee}{\end{equation}}
\newcommand{\bea}{\begin{eqnarray}}
\newcommand{\eea}{\end{eqnarray}}
\newcommand{\bml}{\begin{subequations}}
\newcommand{\eml}{\end{subequations}}
\def\aprle{\buildrel < \over {_{\sim}}}
\def\aprge{\buildrel > \over {_{\sim}}}

\begin{abstract}  
We examine the cosmological evolution equations of de Sitter, flat and anti-de Sitter braneworlds sandwiched in between two n dimensional AdS-Schwarzschild spacetimes. We are careful to use the correct form for the induced Newton's constant on the brane, and show that it would be naive to assume the energy of the bulk spacetime is just given by the sum of the black hole masses. By carefully calculating the energy of the bulk for large mass we show that the induced geometry of the braneworld is just a radiation dominated FRW universe with the radiation coming from a CFT that is dual to the AdS bulk.
\end{abstract}   

\newpage
  
\section{Introduction}

Recent years have seen the development of two very interesting ideas in Theoretical Physics: holography and braneworlds.  For our purposes, a braneworld \cite{RS1,RS2} is an $n-1$ dimensional surface (or brane) marooned in some $n$ dimensional AdS spacetime. In  \cite{RS2}, this extra dimension is infinite. Its geometry is warped exponentially and it is this warp factor that ensures that gravity is localised on the brane.

The notion of holography was given substance by the celebrated AdS/CFT correspondance \cite{Maldacena:adscft,Witten:adscft,Gubser:adscft}. This states that a theory of gravity in a bulk AdS spacetime is dual to a conformal field theory (CFT) on its boundary. Witten \cite{Witten:thermal} argued that if we give a finite temperature to the bulk AdS by considering AdS-Schwarzschild then we find that the CFT on the boundary is also at finite temperature. He found that we could associate the mass, temperature and entropy of the black hole with the corresponding quantities in the boundary CFT.

In \cite{Gubser:gravity}, it was shown that the original Randall-Sundrum braneworlds were equivalent to the AdS/CFT correspondance, with the CFT being cut off in the ultra violet. It is paradoxical to think of a scale invariant theory as having a cut off, so what we actually have is a broken CFT on the brane. As we move the brane towards the boundary we approach the original unbroken version of this duality. Now consider the impact of gravity. When our CFT lives on the boundary, the bulk graviton cannot reach it, and gravity is omitted from the boundary theory. However, for a Randall-Sundrum brane away from the boundary, the bulk graviton \bf can \rm reach the brane and we find that gravity is coupled to the broken CFT.

This model of Randall and Sundrum contains a flat braneworld, where the cosmological constant on the brane is set to zero. We can, however, adjust the brane tension so that we induce a non zero value for the cosmological constant \cite{KR}. Recent observations that our universe may have a small positive cosmological constant suggest that it is these braneworlds that are closer to reality. Such inflationary braneworlds are naturally induced by quantum effects of a field theory on the brane \cite{Reall:BraneNew, Nojiri:braneinflation, Nojiri:entropy} . We shall henceforth refer to flat braneworlds as critical and the dS(AdS) braneworlds as super(sub)critical. As before, we can view these in the context of AdS/CFT \cite{Porrati}. We do indeed find (at least for subcritical walls) that a Karch-Randall compactification is dual to a CFT on the brane, coupled to Einstein gravity.

We wish to examine what happens on these braneworlds when the bulk spacetime is at finite temperature. This occurs naturally for a hot critical braneworld due to the emission of radiation into the bulk \cite{March-Russell:RS2cosmo}. The pure AdS bulk is once again replaced by AdS-Schwarzschild. For a  critical braneworld it was shown \cite{Savonije} that the induced geometry on the brane is exactly that of a standard radiation dominated FRW universe. This radiation is represented by a CFT with an AdS dual description. The issue of AdS/CFT on critical braneworlds embedded in a general class of bulk spacetimes with a negative cosmological constant is discussed in \cite{Shiromizu:ADSnohair}.  In this paper, we attempt to reinforce the AdS/CFT duality by extending the discussion to non-critical braneworlds. There have been previous attempts to do this \cite{Wang}. However, they used a version for the braneworld Newton's constant $G_{n-1}$  in terms of the bulk constant $G_n$, that is only valid for critical walls. We use the correct version \cite{Sasaki, Padilla:instantons} and find that we need to be more careful in calculating the energy of the bulk AdS. In Appendix \ref{naiveappendix} we have shown that we cannot assume the energy is given by the masses of the black holes. By using a procedure similar to that used in \cite{Cappiello}, we properly determine the energy of the bulk. We find that we can similarly describe the braneworld as a radiation dominated FRW universe with the radiation coming from a CFT with an AdS dual.  Our analysis is restricted to $\kappa=1$ closed brane universes, although it would also be interesting to consider $\kappa=0, -1$. Note that the quantum cosmology of braneworlds of arbitrary tension in a pure AdS bulk (as apposed to the AdS-Schwarzschild bulk being investigated here) has been studied from a holographic viewpoint for all values of $\kappa$ \cite{Nunez:stringcosmo, Nunez:quantumcosmo}. 

The rest of this paper is organised as follows: in section 2 we will review the derivation of the equations of motion for the brane embedded in AdS-Schwarzschild. In section 3 we will consider how these equations of motion can be regarded as Friedmann equations for the braneworld. We will properly derive the energy of the bulk spacetime and use this to derive the energy density of the dual CFT. Finally, section 4 contains some concluding remarks.  

\section{The Equations of Motion of the Bulk and the Brane} 

Let us consider two identical $n$ dimensional spacetimes with negative cosmological constant $\Lambda$.  These are glued together by an $n-1$ dimensional braneworld of tension $\sigma$. This is described by the following action:

\begin{equation}
S=\frac{1}{16 \pi G_n}\int_{bulk} d^nx \sqrt{g}(R-2\Lambda) + \frac{1}{8 \pi G_n} \int_{brane} d^{n-1}x \sqrt{h}  2K +\sigma \int_{brane}  d^{n-1}x  \sqrt{h}
\end{equation}
where $g_{ab}$ is the bulk metric and $h_{\mu\nu}$ is the induced metric on the brane. $K_{\mu\nu}$ is the extrinsic curvature of the brane. It appears with a factor of two in the action because we have two copies of the bulk glued together at the brane.

The bulk equation of motion is just given by Einstein's equations with a negative cosmological constant:

\be \label{einstein}
R_{ab} - \half R g_{ab}=-\Lambda g_{ab}
\ee
This admits the following solution \cite{Padilla:nested, Padilla:instantons, Bowcock}:
\be \label{bulk metric}
ds^2_n=-h(Z)dt^2+\frac{dZ^2}{h(Z)}+Z^2d\Omega_{n-2}  
\ee
where 
\be
h(Z)=k_n^2Z^2+1-\frac{c}{Z^{n-3}}
\ee
and $k_n^2$ is related to the bulk cosmological constant by $\Lambda=-\half (n-1)(n-2)k_n^2$. The integration constant $c$ distinguishes between the pure AdS solution ($c=0$) and the AdS-Schwarzschild solution ($c>0$). Meanwhile,  $d\Omega_{n-2}$ is the metric on an $n-2$ sphere.

We now parametrise the brane using the parameter $\tau$. The brane is given by the section $(\bf x\rm^{\mu}, t(\tau), Z(\tau))$ of the general bulk metric. The brane equation of motion is given by the Israel equations for the jump in extrinsic curvature across the brane. We have $Z_2$ symmetry across the brane so these equations just take the following form:

\be \label{israel}
2K_{\mu\nu}-2Kh_{\mu\nu}=-8\pi G_n \sigma h_{\mu\nu}
\ee

The extrinsic curvature is defined in terms of the unit normal, $n^a$, to the brane, by the relation $K_{\mu\nu}=h_{\mu}^ah_{\nu}^b\nabla_{(a}n_{b)}$. Using the fact that:

\be
n_a=(\bf 0 \rm , -\dot{Z}, \dot{t}) \ ; \ \dot{Z}=\frac{dZ}{d\tau} \ ; \ \dot{t}=\frac{dt}{d\tau}
\ee
we arrive at the equation of motion for the brane:
\bea \label{brane EOM}
\dot{Z}^2 &=& aZ^2-1+\frac{c}{Z^{n-3}}   \label{brane EOM1} \\
\ddot{Z} &=& aZ -\left(\frac{n-3}{2}\right) \frac{c}{Z^{n-2}}  \label{brane EOM2} \\
\dot{t} &=& \frac{\sigma_n Z}{h}  \label{brane EOM3}
\eea
where $a=\sigma_n^2-k_n^2$ and $\sigma_n=\frac{4\pi G_n\sigma}{n-2}$.
This analysis has also been presented in  more detail in \cite{Petkou}.

\section{The Cosmology of Non-Critical Braneworlds}

We shall now examine in more detail what is happening on our braneworld when it is sandwiched in between two AdS-Schwarzschild spacetimes. The induced metric is given by the following:

\be \label{induced metric}
ds^2_{n-1}=-d\tau^2+Z(\tau)^2d\Omega_{n-2}
\ee
Notice that the size of our braneworld is given by the radial distance $Z(\tau)$ from the centre of the black hole. Given this structure we can interpret equations (\ref{brane EOM1}) and (\ref{brane EOM2}) as giving rise to the Friedmann equations of the braneworld. If we define the Hubble paramater in the usual way ($H=\frac{\dot{Z}}{Z}$), we arrive at the following equations for the cosmological evolution of the brane:

\bea 
H^2 &=& a-\frac{1}{Z^2}+\frac{c}{Z^{n-1}} \label{cosmo1} \\ 
\dot{H} &=& \frac{1}{Z^2}-\left(\frac{n-1}{2}\right)\frac{c}{Z^{n-1}} \label{cosmo2}
\eea

The cosmological constant term in equation (\ref{cosmo1}) is given by $a$. For $a=0$ we have the critical wall with vanishing cosmological constant. For $a>0$/$a<0$ we have super/subcritical walls that correspond to de Sitter/anti-de Sitter spacetimes. 
As was discussed in \cite{Savonije, Nojiri:entropy}, the brane crosses the black hole horizon  for all values of $a$. For $n>3$, critical walls and subcritical walls have a maximum value of $Z$ only. For supercritical walls there are three possibilities: (i) $Z$ runs from  zero to infinity (or vice versa), (ii) $Z$ runs from infinity down to a strictly positive minimum and then up to infinity again or (iii) $Z$ runs from zero up to a maximum and then down to zero again\footnote{(i) occurs iff $a \geq \left(\frac{n-3}{n-1}\right)\left(\frac{2}{(n-1)c} \right)^{\frac{2}{n-3}}$, otherwise we have (ii) when $Z$ starts out large and (iii) when $Z$ starts out small.}. We will concentrate on the third possibility in this paper, as this is intuitively what one expects from a $\kappa=1$ universe.

The interesting part of equations (\ref{cosmo1}) and (\ref{cosmo2}) lies in the c-term. This is a bulk quantity that should have some interpretation on the brane. The natural interpretation would of course be that it corresponds to the energy density and pressure of a dual CFT. We will consider $c$ to be large so that the contribution of this ``holographic'' term is dominant. We will also restrict ourselves to the region in which our braneworld is near its maximum size. This way we avoid the problems one might encounter near the Big Bang and the Big Crunch, aswell as allowing us to make use of Euclidean quantum gravity, as we shall see.

\subsection{Calculating the energy density of the dual CFT}  

In order to evaluate the energy density of the dual CFT we first need to evaluate the energy of the bulk spacetime. We could naively assume that the energy of bulk is just twice\footnote{``twice'' because we have two copies of AdS-Schwarzschild.} the mass $M$ of the black holes, where the mass is given by the standard formula \cite{Hawking:ADS_blackholes}:

\be \label{BH mass}
M= \frac{(n-2)\Omega_{n-2}c}{16 \pi G_n} 
\ee
and $\Omega_{n-2}$ is the volume of the unit $n-2$ sphere. However, the derivation \cite{Witten:thermal,Hawking:ADS_blackholes} of equation (\ref{BH mass})  includes contributions from the AdS-Schwarzschild spacetime all the way up to the AdS boundary. In our case, we have a brane that has cut off our bulk spacetime before it was able to reach the boundary. We should not therefore include contributions from ``beyond'' the brane and must go back to first principles in order to calculate the energy of the bulk. See Appendix \ref{naiveappendix} to see what happens if we choose the bulk energy to be $2M$.

We will need to Wick rotate to Euclidean signature:

\bea
t &\to& t_E=it \nonumber \\
\tau &\to& \tau_E=i\tau \nonumber
\eea
This is valid provided we restrict ourselves to the region near $\dot{Z}=0$, where we have maximal expansion of the braneworld. By considering $c$ to be large we can guarantee that our Euclidean analysis does not stray away from this region.
Our bulk metric is now given by:

\be \label{Euclidean bulk metric}
ds^2_n=h(Z)dt_E^2+\frac{dZ^2}{h(Z)}+Z^2d\Omega_{n-2}  
\ee
We wish to avoid a conical singularity at the horizon, $Z=Z_H$ where $h(Z_H)=0$. In order to do this we cut the spacetime off at the horizon and associate $t_E$ with $t_E+\beta$ where $\beta=\frac{4\pi}{h^{\prime}(Z_H)}$. The brane is now given by the section $(\bf x \rm^{\mu}, t_E(\tau_E), Z(\tau_E))$ of the Euclidean bulk. The new equations of motion of the brane are the following:

\bea \label{Euclidean brane EOM}
\left(\frac{dZ}{d\tau_E}\right)^2 &=& -aZ^2+1-\frac{c}{Z^{n-3}}   \label{Euclidean brane EOM1} \\
\frac{d^2Z}{d\tau_E^2} &=&-aZ +\left(\frac{n-3}{2}\right) \frac{c}{Z^{n-2}}  \label{Euclidean brane EOM2} \\
\frac{dt_E}{d\tau_E} &=& \frac{\sigma_n Z}{h}  \label{Euclidean brane EOM3}
\eea
It is not difficult to see that for both critical and non critical walls, $Z(\tau_E)$ has a minimum value. In contrast to Lorentzian signature, in Euclidean signature these branes do not cross the black hole horizon. The supercritical wall have a maximum value of $Z$, whilst the critical and subcritical walls may stretch to the AdS boundary. This will not be a problem because the integrand in our overall action will remain finite, as we shall see. 

In calculating the energy we could go ahead and evaluate the Euclidean action of this solution and then differentiate with respect to $\beta$. We must however, remember to take off the contribution from a reference spacetime \cite{Hawking:hamiltonian}. In this context, the most natural choice of the reference spacetime would be pure AdS cut off at a surface, $\Sigma$ whose geometry is the same as our braneworld.

The bulk metric of pure AdS is given by the following:

\be \label{Euclidean pure ADS metric}
ds^2_n= h_0(Z)dT^2+\frac{dZ^2}{h_0(Z)}+Z^2d\Omega_{n-2}
\ee
where
\be
h_0(Z)=k_n^2Z^2+1
\ee
As we said earlier, the cut off surface should have the same geometry as our braneworld. The induced metric on this surface is therefore:

\be \label{surface metric}
ds^2_{n-1}=d\tau_E^2+Z(\tau_E)^2d\Omega_{n-2}
\ee

To achieve this, we must regard our cut off surface as a section  $(\bf x \rm^{\mu}, T(\tau_E), Z(\tau_E))$, where:

\be \label{T condition}
h_0\left(\frac{dT}{d\tau_E}\right)^2+\frac{1}{h_0}\left(\frac{dZ}{d\tau_E}\right)^2=1
\ee

Let us now evaluate the difference $\Delta I$  between the Euclidean action of our AdS-Schwarzschild bulk, $I_{BH}$ and that of our reference background, $I_{AdS}$.

\bea
I_{BH} &=& -\frac{1}{16 \pi G_n}\int_{bulk} d^nx \sqrt{g}(R-2\Lambda) - \frac{1}{8 \pi G_n} \int_{brane} d^{n-1}x \sqrt{h} \ 2K \label{BH action} \\
I_{AdS} &=& -\frac{1}{16 \pi G_n}\int_{ref. \ bulk} d^nx \sqrt{g}(R-2\Lambda) - \frac{1}{8 \pi G_n} \int_{\Sigma} d^{n-1}x \sqrt{h} \ 2K_0 \label{reference action}
\eea
where $K_0$ is the trace of the extrinsic curvature of the cut off surface. Now from equations (\ref{einstein}) and (\ref{israel}), we can immediately obtain:

\bea
R-2\Lambda &=& -2(n-1)k_n^2  \label{ricci}\\
2K &=& 2(n-1) \sigma_n \label{extrinsic}
\eea

The unit normal to the cut off surface, $\Sigma$ is given by $n_a=(\bf 0 \rm, -\frac{dZ}{d\tau_E},  \frac{dT}{d\tau_E})$. We use this to find:
\be
2K_0=(n-1)\frac{2\sigma_n^2Z(\tau_E)+cZ(\tau_E)^{2-n}}{h_0 \frac{dT}{d\tau_E}}
\ee
We will also need the correct form of the measures and the limits in each case. If we say that $-\frac{\beta}{2} \leq t_E \leq \frac{\beta}{2}$, then we obtain the following (see Appendix \ref{appendixa} for a detailed derivation):

\bea
\int_{bulk} d^nx \sqrt{g} \ (R-2\Lambda) &=& 2\Omega_{n-2}\int_{-\frac{\beta}{2}}^{\frac{\beta}{2}} dt_E \frac{Z(\tau_E)^{n-1}-Z_H^{n-1}}{n-1}  \ (R-2\Lambda) \label{action1} \\
\int_{ref. \ bulk} d^nx \sqrt{g}  \ (R-2\Lambda) &=& 2\Omega_{n-2}\int_{-\frac{\beta}{2}}^{\frac{\beta}{2}} dt_E \left(\frac{ \frac{dT}{d\tau_E}}{ \frac{dt_E}{d\tau_E}}\right)\frac{Z(\tau_E)^{n-1}}{n-1}  \ (R-2\Lambda)  \label{action2} \\
\int_{brane} d^{n-1}\sqrt{h} \ 2K &=& \Omega_{n-2} \int_{-\frac{\beta}{2}}^{\frac{\beta}{2}} dt_E  \left(\frac{ 1}{ \frac{dt_E}{d\tau_E}}\right)Z(\tau_E)^{n-2} \ 2K \label{action3} \\
\int_{\Sigma} d^{n-1}\sqrt{h} \ 2K_0 &=& \Omega_{n-2} \int_{-\frac{\beta}{2}}^{\frac{\beta}{2}} dt_E  \left(\frac{ 1}{ \frac{dt_E}{d\tau_E}}\right)Z(\tau_E)^{n-2} \ 2K_0 \label{action4}
\eea
The factor of two in equations (\ref{action1}) and (\ref{action2}) just comes from the fact that we have two copies of the bulk spacetime in each case. Notice that the expressions for the integrals over the brane and the cut off surface $\Sigma$ are the same. This is a consequence of the two surfaces having the same geometry. 
Also using equations (\ref{Euclidean brane EOM3}) and (\ref{T condition}), we put everything together and arrive at the following expression for the difference in the Euclidean action:

\bea
\Delta I &=& \frac{\Omega_{n-2}k_n^2}{4 \pi G_n}  \int_{-\frac{\beta}{2}}^{\frac{\beta}{2}} dt_E Z^{n-1}\left[1-\left(1+\frac{cZ^{1-n}}{\sigma_n^2}\right)^{\half}\left(1-\frac{cZ^{1-n}}{k_n^2}\left(1+\frac{1}{k_n^2Z^2}\right)^{-1}\right) \right] \nonumber \\
& & - \frac{\Omega_{n-2}}{4 \pi G_n} \int_{-\frac{\beta}{2}}^{\frac{\beta}{2}} dt_E \ (n-1)h(Z)Z^{n-3} \left[1-\half \left(1+\frac{cZ^{1-n}}{\sigma_n^2}\right)^{\half}-\half \left(1+\frac{cZ^{1-n}}{\sigma_n^2}\right)^{-\half} \right] \nonumber \\
& & - \frac{\Omega_{n-2}k_n^2}{4 \pi G_n}\beta Z_H^{n-1}
\eea

To proceed further, we are going to have to make things a little bit simpler. As we stated earlier, this analysis is only valid when $c$ is large, and so our bulk is at a very high temperature. By considering this regime we guarantee that we focus on the ``holographic'' energy density, and can ignore contributions from matter on the brane. We have not included any such contributions in our analysis so it is appropriate for us to assume that we are indeed working at large $c$.
To leading order:
\bea
Z_H &\approx& \left(\frac{c}{k_n^2}\right)^{\frac{1}{n-1}} \\
\beta &\approx& \frac{4\pi}{(n-1)k_n^2} \left(\frac{k_n^2}{c}\right)^{\frac{1}{n-1}}
\eea
For supercritical and critical walls we can assume $Z(\tau_E) \gg c^{\frac{1}{n-1}}$. For subcritical walls this is true provided $|a| \ll 1$ (see Appendix \ref{appendixb}). Given this scenario, we now evaluate $\Delta I$ to leading order in $c$:
\be
\Delta I = -\frac{\Omega_{n-2}c\beta}{4 \pi G_n}+\frac{\Omega_{n-2}k_n^2c}{4 \pi G_n}  \int_{-\frac{\beta}{2}}^{\frac{\beta}{2}} dt_E \left(\frac{1}{k_n^2}-\frac{1}{2\sigma_n^2}\right) + \ldots = -\frac{\Omega_{n-2}c\beta}{8 \pi G_n}\left(\frac{k_n^2}{\sigma_n^2}\right)+\ldots
\ee
The entire leading order contribution comes from the bulk rather than the brane, which is consistent with \cite{Witten:thermal}. We can now determine the energy of our bulk spacetime:
\be
E= \frac{d \Delta I} {d \beta} \approx \frac{(n-2)\Omega_{n-2}c}{8\pi G_n}\left(\frac{k_n^2}{\sigma_n^2}\right)
\ee
Notice that in this large $c$ limit, $E \approx 2M\left(\frac{k_n^2}{\sigma_n^2}\right)$, so for critical walls the choice $E=2M$ would indeed have worked.
Our aim was to calculate the energy of the dual CFT, rather than the bulk AdS-Schwarzschild. We must therefore scale $E$, by $\dot{t}$, so that it is measured with respect to the CFT time $\tau$. Recall that we are considering a regime near the maximal expansion of the braneworld. If $Z$ is large,  $\dot{t} \approx \frac{\sigma_n}{k_n^2Z}$ and the energy of the CFT is given by:

\be
E_{CFT}=E\dot{t} \approx \frac{(n-2)\Omega_{n-2}c}{8\pi G_n}\left(\frac{k_n^2}{\sigma_n^2}\right)\left(\frac{\sigma_n}{k_n^2Z}\right)=\frac{(n-2)\Omega_{n-2}c}{8\pi G_n}\left(\frac{1}{\sigma_nZ}\right)
\ee
In order to calculate the energy density we must first evaluate the spatial volume of the CFT:

\be
V_{CFT}=\Omega_{n-2}Z^{n-2}
\ee
We are now ready to give an expression for the energy density of our CFT:

\be
\rho_{CFT}=\frac{E_{CFT}}{V_{CFT}} \approx \frac{(n-2)}{8 \pi G_n \sigma_n}\left( \frac{c}{Z^{n-1}} \right)
\ee

\subsection{The Cosmological Evolution Equations}

Now that we have determined the energy density, $\rho_{CFT}$ of our CFT, we can use the following equation \cite{Savonije} to determine the pressure, $p_{CFT}$:

\be
\dot{\rho}_{CFT}=-(n-2)H(\rho_{CFT}+p_{CFT})
\ee
This yields an expression that is consistent with the CFT corresponding to radiation:

\be \label{pressure}
p_{CFT} \approx \frac{1}{8 \pi G_n \sigma_n}\left(\frac{c}{Z^{n-1}}\right) \approx \frac{\rho_{CFT}}{n-2}
\ee
If we are to make sense of the cosmological evolution equations of the braneworld we will need to know the Newton's constant, $G_{n-1}$ in $n-1$ dimensions. In \cite{Padilla:nested} we proposed that:

\be \label{Newtons constants}
G_{n-1}=\frac{(n-3)}{2}\sigma_n G_n
\ee
This is confirmed by the analysis of \cite{RS2, Sasaki, Shiromizu:ADSnohair, Shiromizu:einstein}.  Using equation (\ref{Newtons constants}) in our expression for $\rho_{CFT}$ gives the more useful expression:

\be
\rho_{CFT} \approx \frac{(n-2)(n-3)}{16 \pi G_{n-1}}\left(\frac{c}{Z^{n-1}}\right)
\ee
We are now ready to insert this and equation (\ref{pressure}) into equations (\ref{cosmo1}) and (\ref{cosmo2}) to derive the cosmological evolution equations for  our braneworld:

\bea 
H^2 &=& a-\frac{1}{Z^2}+\frac{16 \pi G_{n-1}}{(n-2)(n-3)}\rho_{CFT} \label{evolution1} \\ 
\dot{H} &=& \frac{1}{Z^2}- \frac{8 \pi G_{n-1}}{(n-3)}(\rho_{CFT}+p_{CFT})\label{evolution2}
\eea
These are the standard FRW equations in $(n-1)$ dimensions. The braneworld observer therefore sees the normal cosmological expansion driven by the energy density and pressure of the CFT dual to the AdS-Schwarzschild bulk. We have shown this to be true even for non-critical braneworlds, given that we are in near a region of maximal expansion. The conformal field theory behaves like radiation.

\section{Conclusions} \label{section:conclusions}  

In this paper we have examined the cosmological evolution equations for a braneworld sandwiched in between two AdS black holes with large masses. We focused on a region near the maximal expansion of the braneworld and found that the contribution of the black hole energy could be exactly associated with the energy density of a CFT living on the brane. One can regard the evolution of the scale factor as being driven by radiation that is represented by a CFT with an AdS dual.
 
The remarkable thing about this analysis was that it was done in full generality, allowing for all de Sitter, flat, and a large proportion of anti-de Sitter braneworlds. The work of \cite{Savonije} concentrated only on flat braneworlds. Recent observations that we may live in a universe with a small positive cosmological constant suggest that it is important that we extend the discussion at least to de Sitter braneworlds. These have been considered in our paper along with anti-de Sitter braneworlds satisfying $|a| \ll 1$.

Given the mounting evidence for holography in the literature, we are not really surprised by our result. What is interesting is the way in which we were forced to prove it. The proof offered by \cite{Wang} is unacceptable because it relies on the assumption that:

\be
G_{n-1}=\frac{n-3}{2}k_nG_n
\ee
This is true for critical walls, but one should replace $k_n$ in the above expression with $\sigma_n$ when one considers non critical walls \cite{Sasaki,Padilla:instantons}. We also see in Appendix \ref{naiveappendix} that if we had applied the  approach of \cite{Savonije} to non critical walls, a factor of $\frac{k_n^2}{\sigma_n^2}$ would have appeared in front of the CFT terms in equations (\ref{evolution1}) and (\ref{evolution2}). This comes from assuming that the bulk energy is just given by the sum of the black hole masses. As we stated in section 3, this involves an overcounting because it includes energy contributons from ``beyond'' the brane. The correct calculation of the bulk energy given in this paper ensures that the undesirable factor of  $\frac{k_n^2}{\sigma_n^2}$ does not appear.

We need not be restricted to considering the energy density and pressure of the CFT. We could also investigate its other thermodynamic properties. This has been discussed  for flat walls in \cite{Savonije,Cappiello}. One expects that the corresponding results for non critical walls will add even more evidence to the holographic principle. This analysis will be left for future study.

\vskip.5in   
\centerline{\bf Acknowledgements}   
I would like to thank Ian Davies, James Gregory, Ken Lovis, David G. Page, Simon Ross, Clifford Johnson and in particular Ruth Gregory for their helpful discussions. Special thanks also goes to Renata for her continued support. AP was funded by PPARC. 
\medskip

\appendix

\section{Assuming the Bulk Energy is $2M$} \label{naiveappendix}

Here we shall assume that the energy of bulk spacetime is given by:

\be
E=2M
\ee
In order to calculate the energy of the CFT, we should scale $E$ by $\frac{dt}{d\tau_E}$ so that it is measured with respect to the CFT time $\tau$. However, for large $Z$ we have from equation (\ref{brane  EOM3}):

\be
\frac{dt}{d\tau_E}=\frac{\sigma_nZ}{k_n^2Z^2+1-\frac{c}{Z^{n-3}}} \approx  \frac{\sigma_n}{k_n^2Z}
\ee
The energy of the CFT is then:

\be
E_{CFT} = E  \frac{dt}{d\tau_E} \approx 2M \left(  \frac{\sigma_n}{k_n^2Z} \right)
\ee
Since the spatial volume of the CFT is just $V_{CFT}=\Omega_{n-2}Z^{n-2}$, we have the following expression for the energy density of the CFT:

\be
\rho_{CFT} \approx  \frac{2M}{\Omega_{n-2}Z^{n-2}} \left(  \frac{\sigma_n}{k_n^2Z} \right) = \frac{(n-2)}{8 \pi G_n \sigma_n} \left(\frac{c}{Z^{n-1}} \right) \left( \frac{\sigma_n^2}{k_n^2} \right)
\ee
where we have used equation (\ref{BH mass}). We now introduce the correct formula for the Newton's constant in $n-1$ dimensions given by equation (\ref{Newtons constants}). This gives:

\be
\rho_{CFT} \approx \frac{(n-2)(n-3)}{16 \pi G_{n-1}}\left(\frac{c}{Z^{n-1}}\right)\left( \frac{\sigma_n^2}{k_n^2} \right)
\ee
which when inserted back into equation (\ref{cosmo1}) does not in general give the standard form of the Friedmann equation in $n-1$ dimensions:

\be
H^2 = a-\frac{1}{Z^2}+\frac{16 \pi G_{n-1}}{(n-2)(n-3)}\rho_{CFT}\left( \frac{k_n^2}{\sigma_n^2} \right)
\ee
We note that for critical walls the factor of $ \frac{k_n^2}{\sigma_n^2}$ disappears and we do indeed recover the Friedmann equation, although this is not the case for non critical walls.

\section{Limits and Measures for the Action Integrals} \label{appendixa}

Let us consider in more detail each contribution to the action integrals given in equations (\ref{BH action}) and (\ref{reference action}). We will start by looking at the bulk integral for the black hole action:

\be
\int_{bulk}=\int_{bulk} d^n x \sqrt{g}(R-2\Lambda)
\ee
From equation (\ref{ricci}), we see that $R-2\Lambda$ is constant and so does not cause us any problems. Given that the AdS-Schwarzschild bulk is cut off at the brane, $Z(\tau_E)$, and the horizon, $Z_H$,  we find that:

\be
\int_{bulk}=2\Omega_{n-2}\int_{-\frac{\beta}{2}}^{\frac{\beta}{2}} dt_E \int_{Z_H}^{Z(\tau_E)} dZ \ Z^{n-2} (R-2\Lambda)=2\Omega_{n-2}\int_{-\frac{\beta}{2}}^{\frac{\beta}{2}} \frac{Z(\tau_E)^{n-1}-Z_H^{n-1}}{n-1} (R-2\Lambda)
\ee
which is just equation (\ref{action1}). The factor of two comes in because we have two copies of AdS-Schwarzschild. The factor of $\Omega_{n-2}$ just comes from integrating out $\int d\Omega_{n-2}$. We now turn our attention to the bulk integral for the reference action:

\be
\int_{ref. \ bulk}=\int_{ref. \ bulk} d^n x \sqrt{g}(R-2\Lambda)
\ee
Again, $R-2\Lambda$ is constant and does not worry us. This time the AdS bulk is cut off at $\Sigma$ (given by $Z=Z(\tau_E)$), and at $Z=0$. The periodicity of the $T$ coordinate is $\beta^{\prime}$ rather than $\beta$. The bulk integral for the reference action is then:

\be
\int_{ref. \ bulk}=2\Omega_{n-2}\int_{-\frac{\beta^{\prime}}{2}}^{\frac{\beta^{\prime}}{2}} dT \int_{0}^{Z(\tau_E)} dZ \ Z^{n-2} (R-2\Lambda) =2\Omega_{n-2}\int_{-\frac{\beta^{\prime}}{2}}^{\frac{\beta^{\prime}}{2}} \frac{Z(\tau_E)^{n-1}}{n-1} (R-2\Lambda)
\ee
$\beta^{\prime}$ is fixed by the condition that the geometry of $\Sigma$ and the brane should be the same. This just amounts to saying that $T^{-1}(\pm \frac{\beta^{\prime}}{2})=\pm \tau_{max}=t_E^{-1}(\pm \frac{\beta}{2})$ where $-\tau_{max} \leq \tau_E \leq \tau_{max}$ on both $\Sigma$ and the brane. As illustrated below by changing coordinates to $\tau_E$ and then $t_E$, we arrive at equation (\ref{action2}):

\bea
\int_{ref. \ bulk} &=& 2\Omega_{n-2}\int_{-\tau_{max}}^{\tau_{max}} d\tau_E \ \frac{dT}{d\tau_E} \frac{Z(\tau_E)^{n-1}}{n-1} (R-2\Lambda) \nonumber \\
&=& 2\Omega_{n-2}\int_{-\frac{\beta}{2}}^{\frac{\beta}{2}} dt_E \frac{d\tau_E}{dt_E} \frac{dT}{d\tau_E} \frac{Z(\tau_E)^{n-1}}{n-1} (R-2\Lambda)
\eea
Consider now the brane integral:
\be
\int_{brane}= \int_{brane} d^{n-1}x \sqrt{h}\ 2K
\ee
We will use the coordinate $\tau_E$ to begin with and then change to $t_E$, thus arriving at equation (\ref{action3}):
\be 
\int_{brane}=\Omega_{n-2}\int_{-\tau_{max}}^{\tau_{max}}d\tau_E \ Z(\tau_E)^{n-2} \ 2K =\Omega_{n-2}\int_{-\frac{\beta}{2}}^{\frac{\beta}{2}} dt_E \frac{d\tau_E}{dt_E} Z(\tau_E)^{n-2} \ 2K
\ee
The procedure for arriving at equation (\ref{action4}) is exactly the same, owing to the fact that $\Sigma$ and the brane have the same geometry.

\section{Justifying $Z(\tau_E) \gg c^{\frac{1}{n-1}}$ in large $c$ limit} \label{appendixb}

Let us consider the claim made in section 3.1 that for most brane solutions,  $Z(\tau_E) \gg c^{\frac{1}{n-1}}$ in the large $c$ limit. The governing equation for the branes in Euclidean AdS-Schwarzschild is given by equation (\ref{Euclidean brane EOM1}):

\be 
\left(\frac{dZ}{d\tau_E}\right)^2 = -aZ^2+1-\frac{c}{Z^{n-3}}  
\ee 
Now in each case, $Z \geq Z_{min}$ where $Z_{min}$ is the minimum value of $Z$ on the brane. It is sufficient to show that $Z_{min} \gg c^{\frac{1}{n-1}}$. At $Z=Z_{min}$, $\frac{dZ}{d\tau_E}=0$.
For $a=0$,  we have:

\be
Z_{min}=c^{\frac{1}{n-3}} \gg c^{\frac{1}{n-1}}
\ee
For $a>0$, we have:

\be 
Z_{min} \geq c^{\frac{1}{n-3}} \gg c^{\frac{1}{n-1}}
\ee
We see that our claim holds for supercritical and critical walls. For subcritical walls with $a<0$ we need to be more careful. $Z_{min}$ satisfies:

\be
Z_{min}^{n-3}(1+|a|Z_{min}^2)=c
\ee
If $Z_{min}^2 \ll |a|^{-1}$ then $Z_{min} \approx c^{\frac{1}{n-3}}$. If $Z_{min}^2 \sim |a|^{-1}$ then $(1+|a|Z_{min}^2) \sim c^0$ and therefore $Z_{min} \sim  c^{\frac{1}{n-3}}$. In each case we have $Z_{min} \gg  c^{\frac{1}{n-1}}$.
Finally, when $Z_{min}^2 \gg |a|^{-1}$:

\be
Z_{min} \approx \left(\frac{c}{|a|}\right)^{ \frac{1}{n-1}}
\ee 
Provided $|a| \ll 1$ we can say:
\be
Z_{min} \gg c^{ \frac{1}{n-1}}
\ee
We see, therefore that the claim made in section 3.1 was indeed valid: $Z(\tau_E) \gg c^{\frac{1}{n-1}}$ for subcritical walls with $|a| \ll 1$ and  all supercritical and  critical walls.

\bibliographystyle{utphys}

\bibliography{tony}

\end{document}